# Wavelike fracture pattern in metallic glasses: a Kelvin-Helmholtz flow instability

M.Q. JIANG,[1,2 (a)] G. WILDE,[2] C. B. QU,[3] F. JIANG,[4] H. M. XIAO,[3] J. H. CHEN,[1] S. Y. FU,[3] L.H. DAI[1 (b)]

[1]*State Key Laboratory of Nonlinear Mechanics, Institute of Mechanics, Chinese Academy of Sciences, Beijing 100190, China*

[2]*Institute of Materials Physics, Westfälische Wilhelms-Universität Münster, Münster 48149, Germany*

[3]*Technical Institute of Physics and Chemistry, Chinese Academy of Sciences, Beijing 100190, China*

[4]*State Key Laboratory for Mechanical Behavior of Materials, Xi'an Jiaotong University, Xi'an 710049, China*

PACS 64.70.pe－Metallic glasses

PACS 62.20.mm－Fracture

PACS 47.20.Ft－Instability of shear flows (e.g., Kelvin-Helmholtz)

**Abstract**－We report a wavelike fracture pattern in a Zr-based bulk metallic glass that has been deformed under quasi-static uniaxial tensions at temperatures between room temperature (300 K) and liquid nitrogen temperature (77 K). We attribute this wavelike pattern to a Kelvin-Helmholtz flow instability that occurred at certain

---

(a) E-mail: mqjiang@imech.ac.cn
(b) E-mail: lhdai@lnm.imech.ac.cn

interfaces between local cracking/softening regions. The instability criterion for the pattern formation is achieved via a hydrodynamic perturbation analysis, and furthermore an instability map is built which demonstrates that the shear velocity difference on both sides of the interface is the main destabilizing factor. Finally, the characteristic instability time (the inverse of the instability growth rate) is explored by seeking the dispersion relation in the dominant (fastest) instability mode. The results increase the understanding of the flow and fracture of metallic glasses as well as the nature of their liquid structures.

One of the most striking features of a fluid is the ability to flow due to a very limited resistance to shear. When a fluid is subjected to a significant shear, its flow is prone to be mechanically unstable into a turbulence where flow instabilities are frequently manifested by the emergence of waves or vortices [1-3]. Nevertheless, flow instabilities are not exclusive features of fluids, and actually can also take place in solids under the activation of sufficiently high energy. For example, confined sliding [4], high-speed impact [5] or nanosecond pulse-laser [6] can induce the formation of (primitive) vortical structures in the plastic flow of crystalline metals, indicating the occurrence of flow instabilities. This situation is of paramount significance to metallic glasses, since this type of disordered solids, in the popular and basically correct conception, are "frozen" liquids that have lost their ability to flow at the glass transition temperature [7-8]. In metallic glasses, a widely reported flow instability is the Saffman-Taylor (S-T) flow instability, which describes that a crack (inviscid air) is pushed into a more viscous liquid layer (a shear band) driven by a negative pressure gradient [9-10]. The viscous fingering recently observed in nanosecond pulse-laser ablating a Zr-based bulk metallic glass also points toward the S-T flow instability [11]. Naturally one would like to ask: are there any other types of flow instabilities that can take place in metallic glasses? In the present letter, we observed some wavelike configurations on the fracture surfaces of a typical Zr-based (Vitreloy 1) bulk metallic glass subjected to quasi-static uniaxial tensions over a wide range of temperatures. It is proposed that the wavelike fracture pattern results from the Kelvin-Helmholtz (K-H) flow instability that is governed by the competition between a shear velocity gradient and the surface tension.

A bulk metallic glass with the composition of $Zr_{41.2}Ti_{13.8}Cu_{12.5}Ni_{10.0}Be_{22.5}$ (Vitreloy 1) was adopted in this study because of its excellent glass-forming ability and high thermal stability compared with other systems [12]. The as-cast plates were machined into dogbone-like specimens with the gauge dimension of $13\ mm \times 2\ mm \times 2\ mm$ for tensile tests. Uniaxial tension tests were performed using an Instron material testing machine under a strain rate of $\sim 10^{-4}\ s^{-1}$ at the following temperatures: room temperature (300 K), 221 K, 152 K and liquid nitrogen temperature (77 K). In all cases, the Vitreloy 1 glass displays an elastic-brittle failure without macroscopically appreciable plasticity. The fracture angles fall in the range of $55^0$-$61^0$, demonstrating that the fracture is dominated by shear stress but affected by tensile normal stress [13-15]. After testing, an FEI Sirion scanning electron microscope was used to examine the fracture surfaces of all specimens.

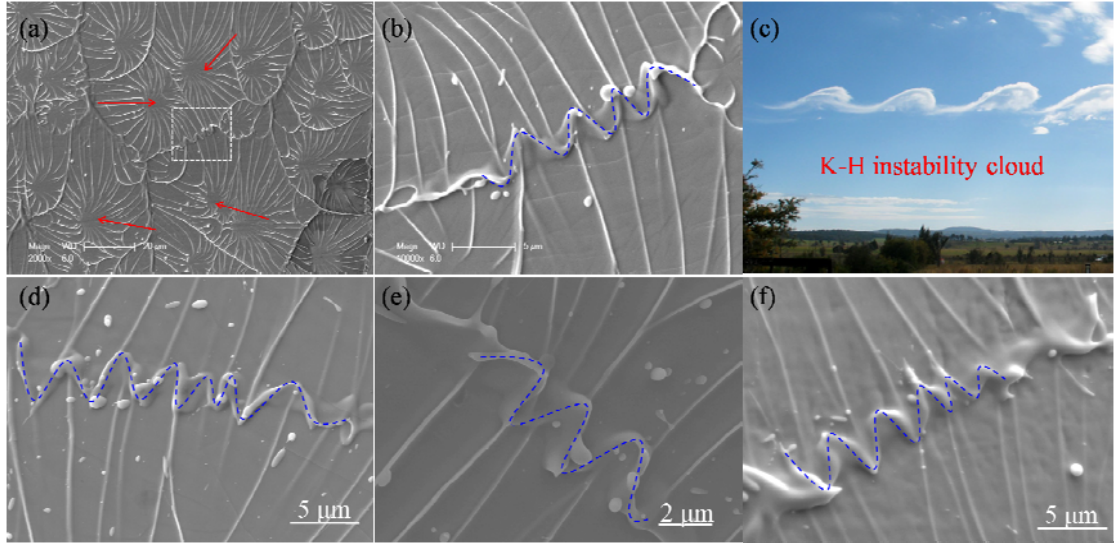

Fig. 1: (Color online) Microscale wavelike structures on tensile fracture surfaces of Vitreloy 1 metallic glasses at temperatures of 300 K (a and b), 221 K (d), 152 K (e) and 77 K (f). (b) is the magnified views of the rectangular area in (a). (c) Similarly shaped K-H instability cloud (Ref. 20) on a scale $10^8$ times that of (b, d-f).

Figure 1(a) shows the representative fracture morphology of the Vitreloy 1 at 300 K. It is noted that the fracture surface consists of many sub-regions with different sizes. In each sub-region, there always exists a smooth core [marked by the arrows in fig. 1(a)] and radiating ridges of veins point to the core. It has been accepted that these smooth cores should be the nucleation sites of local sub-cracks within the shear band prior to the final fracture [13, 16-18], and the vein ridges are ascribed to the S-T instability at the interface between the advancing sub-crack and the liquid layer ahead of the sub-crack tip [9-10]. Quite interestingly, at some boundaries between the sub-regions, wavelike patterns on a microscale are observed, as shown in fig. 1(b) which displays an enlarged view of the rectangular area in fig. 1(a). This kind of interface configuration directs our attention to the K-H instability in hydrodynamics [19]. For instance, fig. 1(c) shows a typical K-H instability case [20], a "billow cloud", that is very rare and exceptional but people can see it some days. The K-H instability cloud is the result of an interface instability between the cloud and the wind due to strong relative shear. The wavelike fracture pattern in fig. 1(b) is very similar to the "billow cloud" in fig. 1(c), although their scales are different by at least eight orders of magnitude. More importantly, the wavelike configurations can be also observed in other temperature cases: 221 K, 152 K and 77 K, as displayed in figs. 1(d)-1(f), respectively. It is noticed that the characteristic wavelength (about 1-5 μm) of these configurations are almost not affected by the temperature change, which implies that their formation is mainly determined by the flow dynamics. The temperature-independence of fracture patterns is also consistent with the fact that shear bands prior to fractures are in the flow state with a saturation concentration of plastic flow-induced free volume [21].

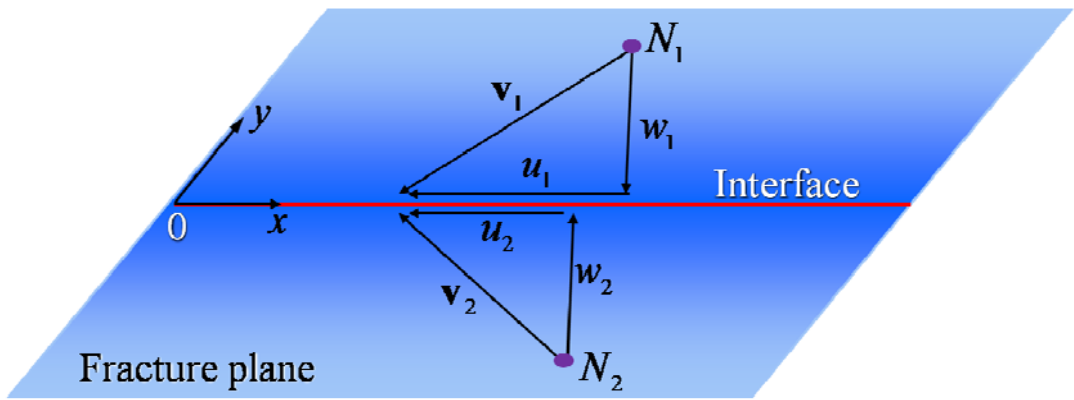

Fig. 2: (Color online) Schematics of the interplay of two sub-cracks nucleated at the sites $N_1$ and $N_2$, respectively, on the main fracture plane modeled in an Eulerian coordinate system $(x, y)$. The two sub-cracks propagate with the velocities of $\mathbf{v}_1$

and $\mathbf{v}_2$, respectively.

Obviously, the wavelike pattern at the interface should result from the interplay of local sub-cracks. Without loss of generality, we consider two nucleation sites, $N_1$ and $N_2$, of the local sub-cracks on the main fracture surface that is localized in an Eulerian coordinate system ($x, y$), as illustrated in fig. 2. Once nucleated, the two sub-cracks begin to advance with the velocities of $\mathbf{v}_1$ and $\mathbf{v}_2$, respectively. The advancing sub-crack pushes the plastic zone ahead of the crack tip with the same velocity. There is general consensus that the material (shear band) in the plastic zone at the crack tip can be considered as a viscous liquid [15, 22-25]. Therefore, the propagation of the sub-cracks should inevitably lead to a relative shear-motion of two viscous liquids in the vicinity of their interface that is assumed to be at $y = 0$. A clear evidence for the shear motion at the interfaces is from the observation that radiating ridges of veins due to S-T instability is not strictly perpendicular to the interfaces [see fig. 1]. This leads to a difference in the velocities of two neighboring sub-cracks or fluids projecting on their interface. It is reasonable to believe that this relative motion of the two fluids induces the K-H flow instability that manifests in the form of a wavelike interface. Here we assume the two fluids as incompressible, and the effect of surface tension between them is taken into account. The hydrodynamic equations for the two fluids and their interface can be written as

$$\rho D_t u + \partial_x p - \mu \Delta u = 0, \tag{1}$$

$$\rho D_t w + \partial_x p - \mu \Delta w = \sigma \frac{\partial^2 y_s}{\partial x^2} \delta(y - y_{s0}), \tag{2}$$

$$D_t \rho = 0, \tag{3}$$

$$\nabla \cdot \mathbf{v} = 0, \tag{4}$$

$$D_t y_s = w_s, \tag{5}$$

where (1) and (2) are the conservation laws for momentum along the $x$-direction and $y$-direction, respectively, (3) the incompressibility condition, (4) the continuity equation and (5) the interface continuity condition. In these equations, $D_t \equiv \partial_t + \mathbf{v} \cdot \nabla$ is the material derivation, $\rho$ is the density, $u$ and $w$ are the components of $\mathbf{v}$ along the $x$-direction and $y$-direction, respectively, $\mu$ denotes the coefficient of viscosity, $p$ is the pressure, $\sigma$ is the surface tension constant, $\delta$ is the Dirac delta function, the subscript "$s$" denotes the interface, and $y_{s0} = 0$ is the initial position of the interface.

We consider a small deviation $y_s'$ of the initial interface ($y_{s0}$) with the form $\exp i(\omega t + kx)$. This perturbation will cause small deviations $\{u', w', p', \rho', \mu'\}$ of other hydrodynamic variables $\{u_0, w_0, p_0, \rho_0, \mu_0\}$ with an identical form. It is therefore interesting to concentrate on the evolution of the disturbed interface. In order to highlight the essential physics, we assume that $u_0 = u_0(y)$, $w_0 = \text{const.1}$, $\rho_0 = \rho_0(y)$, and $\mu_0 = \text{const.2}$. Here, $k$ is the wave number related to the spatial size of the instability and $\omega = \gamma + i\alpha$, here, $\alpha$ is the growth rate of the instability. The stability of the interface is actually determined by the sign of $\alpha$. If $\alpha \leq 0$, it is stable, otherwise, unstable.

Inserting the perturbations into Eqs. (1)-(5) and only retaining the first order terms of the perturbed velocity $w'$, we can derive the governing equation for the perturbed state:

$$\begin{aligned}
&-i\rho_0(\omega+ku_0)(\Gamma^2-k^2)w' + \mu_0(\Gamma^2-k^2)\Gamma^2 w' - \mu_0(\Gamma^2-k^2)k^2 w' \\
&-i\Gamma\rho_0(\omega+ku_0)\Gamma w' - ik\rho_0\Gamma u_0\Gamma w' \\
&+ik\Gamma\rho_0\Gamma u_0 w' + ik\rho_0\Gamma^2 u_0 w' + ik\rho_0\Gamma u_0\Gamma w' - ik\Gamma\mu'\Gamma^2 u_0 - ik\mu'\Gamma^3 u_0 \\
&= -\sigma k^4 y'\delta(y)
\end{aligned} \quad (6)$$

where $\Gamma = \partial_y$. Through the integral approximation, the interface condition that is satisfied at a surface of the discontinuity can be given by:

$$\Delta_s\left[-\rho_0 i(\omega+ku_0)\Gamma w' + i\rho_0 k\Gamma u_0 w' + \mu_0(\Gamma^2-k^2)\Gamma w' - i\mu' k\Gamma^2 u_0 - \mu_0 k^2\Gamma w'\right] = -\sigma k^4 y'_s, \quad (7)$$

where $\Delta_s = f_{y_{s0}+0} - f_{y_{s0}-0}$. It should be kept in mind that the domain of interest is the two thin fluid layers in the vicinity of the interface. Thus we shall further suppose that the two fluids are flowing with the constant velocities $u_{01}$ and $u_{02}$ as well as at constant densities $\rho_{01}$ and $\rho_{02}$. We can immediately obtain the solution of Eq. (6) except for the interface, that is

$$w'_n = y'_s i(\omega+ku_{01})e^{(-1)^n ky} \quad (n=1, 2). \quad (8)$$

Applying the solution to the interface condition (7) yields the spectral equation for the growth rate $\alpha$ of the perturbation:

$$a_4\alpha^4 + a_3\alpha^3 + a_2\alpha^2 + a_1\alpha + a_0 = 0, \quad (9)$$

with

$$a_4 = 4(\rho_{01}+\rho_{02})^3, \quad (10a)$$

$$a_3 = -8k^2 (\mu_{01} + \mu_{02})(\rho_{01} + \rho_{02})^2, \tag{10b}$$

$$a_2 = k(\rho_{01} + \rho_{02})[5k^3(\mu_{01} + \mu_{02})^2 + 4k^2\sigma(\rho_{01} + \rho_{02}) - 4k\rho_{01}\rho_{02}(u_{01} - u_{02})^2], \tag{10c}$$

$$a_1 = -k^3(\mu_{01} + \mu_{02})[k^3(\mu_{01} + \mu_{02})^2 + 4k^2\sigma(\rho_{01} + \rho_{02}) - 4k\rho_{01}\rho_{02}(u_{01} - u_{02})^2], \tag{10d}$$

$$a_0 = k^7\sigma(\mu_{01} + \mu_{02})^2 - k^6(u_{01} - u_{02})^2(\rho_{01}\mu_{02}^2 + \rho_{02}\mu_{01}^2). \tag{10e}$$

According to the Routh-Hurwitz criterion [26], the instability criterion for the interface can be obtained as

$$|u_{01} - u_{02}| > (\mu_{01} + \mu_{02})\sqrt{\frac{\sigma k}{\rho_{01}\mu_{02}^2 + \rho_{02}\mu_{01}^2}}. \tag{11}$$

This criterion indicates that the difference in initial horizontal velocity $\Delta u_0 = |u_{01} - u_{02}|$ incurs the interface instability, whereas the surface tension $\sigma$ suppresses the instability development. In physics, this kind of configuration just satisfies the K-H instability [19].

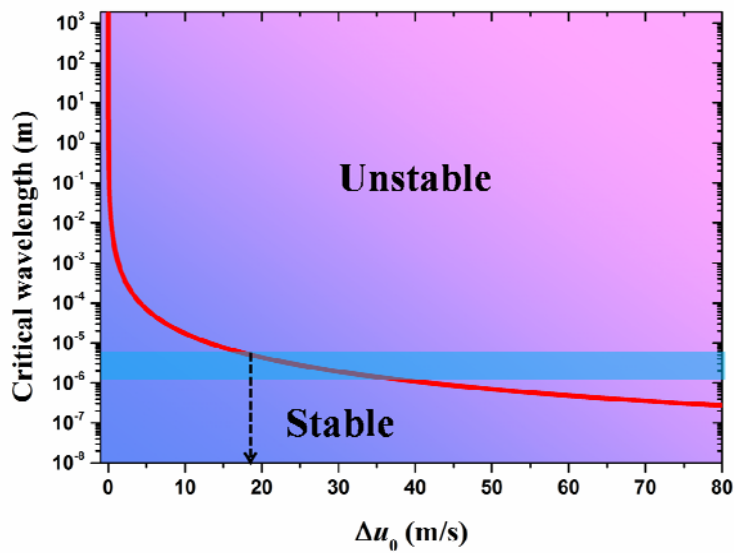

Fig. 3: (Color online) An instability map constructed by the plot of the critical

instability wavelength as a function of the relative shear velocity of two viscous fluids.

Based on the instability criterion (11), the dynamic balance between the stabilizing and destabilizing effects determines a critical wavelength:

$$\lambda_c = \frac{2\pi(\mu_{01}+\mu_{02})^2}{\rho_{01}\mu_{02}^2 + \rho_{02}\mu_{01}^2} \frac{\sigma}{\Delta u_0^2}. \tag{12}$$

The critical wavelength, from a viewpoint of the spatial size, determines whether a perturbation will grow unstably or die out. Perturbations with a wavelength smaller than $\lambda_c$ will disappear, whereas the ones with a wavelength larger than $\lambda_c$ will lead to a runaway instability. Equation (12) shows that the critical instability wavelength depends not only on the material parameters ($\rho_0$, $\mu_0$ and $\sigma$) but also on the relative shear velocity $\Delta u_0$ of the two fluids. According to our previous work [27], the crack propagation velocity of the Vitreloy 1 is about 1%-2% of the Rayleigh wave speed (about 2200 ms$^{-1}$) at the vein-pattern stage. Thus, an interval $0 \leq \Delta u_0 \leq 80$ ms$^{-1}$ is reasonably obtained. Furthermore, we neglect the trivial difference in density and viscosity of the two fluids, that is, $\rho_{01} \approx \rho_{02} = 6020$ kg/m$^3$ [28], and $\mu_{01} \approx \mu_{02} = 10^5$ Pas that is comparable to the material viscosity within shear bands [23, 29-30]. We plot the critical instability wavelength versus the relative shear velocity for the surface tension $\sigma \approx 0.83$ Nm$^{-1}$ [31], which constructs an instability map (see fig. 3). The map predicts that the critical instability wavelength decreases with increasing $\Delta u_0$, which separates the domain into an unstable (upper) and a stable (lower) one. Higher $\Delta u_0$ will make the instability take place easier. At the limit $\Delta u_0 \to 0$, it predicts $\lambda_c \to \infty$, which means a stable state. For the present experiments, the characteristic wavelength of the wavelike patterns is about 1-5 μm (fig. 1) that is marked by a blue band in the instability map (fig. 3). According to the map, the relative shear velocity of the two fluids should be larger than a critical value (about 20 ms$^{-1}$) to guarantee the occurrence of the K-H instability. This explains why the wavelike pattern can only be observed at certain interfaces that just satisfy the K-H instability condition, i.e., Eq. (11) or (12), instead of observing it at all interfaces. Furthermore, the good agreement between the predicted critical wavelength and the measured wavelength validates the K-H instability mechanism for the observed wavelike fracture patterns.

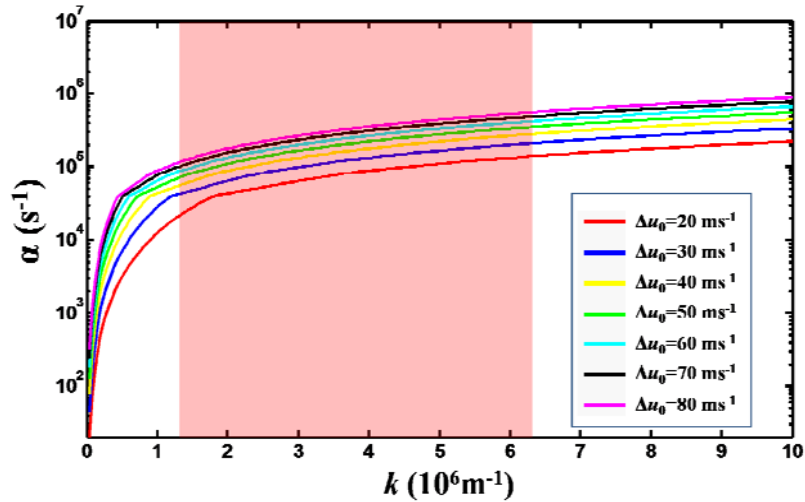

Fig. 4: (Color online) The dispersion relations in the dominant instability mode at different relative shear velocities. The red region corresponds to the current observation.

It is also interesting to examine the characteristic instability time to see if there is a sufficient time to allow the K-H instability to occur. The characteristic instability time is the inverse of the instability growth rate in the dominant (fastest) mode that is determined by the dispersion relation $\partial \alpha_m / \partial k_m = 0$. We plotted it for $\Delta u_0 \geq 20$ ms$^{-1}$ in fig. 4. It can be seen that for each $\Delta u_0$ the instability growth rate increases with increasing wave number. This means that the smaller the instability wavelength, the faster the instability takes place. Furthermore, for a fixed wave number (or instability wavelength), a higher $\Delta u_0$ will increase the instability growth rate, also leading to a faster instability. According to our current observations, the characteristic wavelength (about 1-5 μm) of the wavelike patterns corresponds to the range $1.26 \times 10^6$ m$^{-1}$ $\leq k \leq 6.28 \times 10^6$ m$^{-1}$ indicated by a red region in fig. 4. In this range of wave numbers, the characteristic instability time ranges from microseconds to tens of microseconds that is less than the characteristic time of shear banding (inhomogeneous flow) propagation [32-33]. This ensures that the K-H instability has sufficient time to occur during inhomogeneous flow in front of moving cracks. However, the observed wavelike configurations induced by the K-H instability do not develop into mature vortices. The reason is maybe related to the highly dissipative nature of metal plasticity and the greater resistance to flow, which prevents the further development of the wavelike structures and confines them to the early stage of the onset of turbulence.

In summary, quasi-static uniaxial tensile experiments were performed on Vitreloy 1 bulk metallic glasses between room temperature and liquid nitrogen temperature. In this temperature range, wavelike patterns are observed on the facture surfaces. A hydrodynamic instability analysis reveals that this wavelike fracture pattern results from the K-H flow instability of viscous fluids at sub-crack tips. Such mechanism is highlighted by exploring the critical instability wavelength and the characteristic instability time. We believe that wavelike flow patterns in metallic glasses could be observed in other situations [34], so long as the underlying K-H instability mechanism becomes activated.

* * *

Financial support is from the NSFC (Grants Nos. 11372315, 11132011 and 11023001). M.Q.J. acknowledges the Alexander von Humboldt Foundation for support with a post-doctoral fellowship. This work is also partially supported by the State Key Laboratory for Mechanical Behavior of Materials, Xi'an Jiaotong University.